\documentclass[10pt,twocolumn, aps,prl,showpacs,superscriptaddress]{revtex4}
\usepackage{graphicx, amssymb, amsmath, dsfont}
\usepackage{subfigure}
\usepackage{color}
\definecolor{DarkGray}{rgb}{0.1,0.1,0.5}
\usepackage{url}
\usepackage[colorlinks=true,breaklinks, linkcolor= DarkGray,citecolor= DarkGray,urlcolor= DarkGray]{hyperref}	
\newcommand{\eqnref}[1]{\hyperref[#1]{{(\ref*{#1})}}}
\newcommand{\thmref}[1]{\hyperref[#1]{{Theorem~\ref*{#1}}}}
\newcommand{\lemref}[1]{\hyperref[#1]{{Lemma~\ref*{#1}}}}
\newcommand{\corref}[1]{\hyperref[#1]{{Corollary~\ref*{#1}}}}
\newcommand{\defref}[1]{\hyperref[#1]{{Definition~\ref*{#1}}}}
\newcommand{\secref}[1]{\hyperref[#1]{{Section~\ref*{#1}}}}
\newcommand{\figref}[1]{\hyperref[#1]{{Fig.~\ref*{#1}}}}
\newcommand{\tabref}[1]{\hyperref[#1]{{Table~\ref*{#1}}}}
\newcommand{\remref}[1]{\hyperref[#1]{{Remark~\ref*{#1}}}}
\newcommand{\appref}[1]{\hyperref[#1]{{Appendix~\ref*{#1}}}}
\newcommand{\claimref}[1]{\hyperref[#1]{{Claim~\ref*{#1}}}}
\newcommand{\exampleref}[1]{\hyperref[#1]{{Example~\ref*{#1}}}}
\newcommand{\comment}[1]{\emph{\color{blue}Comment:\color{black} #1}} 
 
\newlength{\commentslength}
\newcommand{\comments}[1]{
\hspace{-2\parindent}
\addtolength{\commentslength}{-\commentslength}
\addtolength{\commentslength}{\linewidth}
\addtolength{\commentslength}{-\parindent}
\fcolorbox{blue}{white}{\smallskip\begin{minipage}[c]{\commentslength}
\emph{Comments:}\begin{itemize}#1\end{itemize}\end{minipage}}\bigskip
}
\renewcommand{\comment}[1]{}\renewcommand{\comments}[1]{}

\newcommand{\ket}[1]{\left| #1\right\rangle}      
\newcommand{\bra}[1]{\left\langle #1\right|}

\newcommand{\abs}[1]{\left|#1 \right|}			
\newcommand{\N}{\mathbb{N}}
\newcommand*{\cB}{\mathcal{B}}
\newcommand{\cC}{{\mathcal C}}
\newcommand{\cD}{{\mathcal D}}	
\newcommand{\cH}{{\mathcal H}}

\DeclareMathOperator{\TV}{\operatorname{TV}}
\DeclareMathOperator{\WRT}{\operatorname{WRT}}

\DeclareMathOperator{\MCG}{\operatorname{MCG}}
\DeclareMathOperator{\SU}{\operatorname{SU}}
\DeclareMathOperator{\SO}{\operatorname{SO}}
\DeclareMathOperator{\GL}{\operatorname{GL}}

\newtheorem{theorem}{Theorem}

\def\vac {0}

\begin{document}

\title{Approximating Turaev-Viro 3-manifold invariants \\is universal for quantum computation}

\author{Gorjan Alagic}
\affiliation{Institute for Quantum Computing, University of Waterloo}
\author{Stephen P. Jordan}
\affiliation{Institute for Quantum Information, California Institute of Technology}
\author{Robert K\"onig}
\affiliation{Institute for Quantum Information, California Institute of Technology}
\author{Ben W. Reichardt}
\affiliation{Institute for Quantum Computing, University of Waterloo}

\date{\today}

\begin{abstract}
The Turaev-Viro invariants are scalar topological invariants of compact, orientable $3$-manifolds.  
We give a quantum algorithm for additively approximating Turaev-Viro invariants of a manifold presented by a Heegaard splitting.  The algorithm is motivated by the relationship between topological quantum computers and $(2+1)$-D topological quantum field theories.  Its accuracy is shown to be nontrivial, as the same algorithm, after efficient classical preprocessing, can solve any problem efficiently decidable by a quantum computer.  Thus approximating certain Turaev-Viro invariants of manifolds presented by Heegaard splittings is a universal problem for quantum computation.  This establishes a novel relation between the task of distinguishing non-homeomorphic $3$-manifolds and the power of a general quantum computer.  
\end{abstract}


\maketitle

The topological quantum computer is among the most striking examples of known relationships between topology and physics.  In such a computer, quantum information is encoded in a quantum medium on a $2$-D surface, whose topology determines the ground space degeneracy.  Surface deformations implement encoded operations.  Topological quantum computers are universal, i.e., can implement arbitrary quantum circuits.  It is natural to try to identify the topological origin of this computational power.  

One answer is that the power stems from the underlying $(2+1)$-D topological quantum field theory (TQFT)~\cite{FreedmanKitaevWang00}.  The TQFT assigns a Hilbert space $\cH_\Sigma$ to a $2$-D surface $\Sigma$, and a unitary map $U(f): \cH_\Sigma \rightarrow \cH_{\Sigma'}$ to every diffeomorphism $f: \Sigma \rightarrow \Sigma'$, subject to a number of axioms~\cite{Walker91}.  However, this answer is not fully satisfactory; the definition of a TQFT is involved, and uses mathematics that appears in similar form in the theory of quantum computation.  A second answer, arising in~\cite{AharonovJonesLandau06Jones, AharonovArad06Jones, GarneroneMarzuoliRasetti06Jones, WocjanYard07Jones}, is that quantum computers' power comes from their ability to approximate the evaluation, at certain points, of the Jones polynomial of the plat closure of a braid.  

Here we give an alternative topological description of the power of quantum computers, in terms of the Turaev-Viro $3$-manifold invariants.  Observe that restricting TQFTs to closed manifolds results in scalar invariants.  We show that approximating certain such invariants is equivalent to performing general quantum computations.  That is, we give an efficient quantum algorithm for additively approximating Turaev-Viro invariants, and conversely we show that for any problem decidable in bounded-error, quantum polynomial time (BQP), there is an efficient classical reduction to the Turaev-Viro invariant approximation problem.  The classical procedure outputs the description of a $3$-manifold whose certain Turaev-Viro invariant is either large or small depending on whether the original BQP algorithm outputs $1$ or $0$.  

Turaev and Viro~\cite{TuraevViro92} defined a family of invariants for compact, orientable $3$-manifolds.  The original definition parameterized the invariants by the quantum groups $\SU(2)_k$, for $k \in \N$, but it was extended by Barrett and Westbury~\cite{BarrettWestbury96invariants} to give an invariant for any spherical tensor category~$\cC$.  Any compact $3$-manifold $M$ is homeomorphic to a finite collection of tetrahedra glued along their faces~\cite{Moise52}.  Beginning with such a triangulation, assign a certain rank-six tensor $F$ to each tetrahedron and a certain gluing tensor $d$ to every edge.  The invariant $\TV_\cC(M)$ is  
the contraction of the tensor network, which can be written out as 
\begin{equation} \label{eq:triang_invar}
\hspace{-0.5ex} \TV_\cC(M) = 
\cD^{-2 \abs V}\hspace{-0.2ex} \sum_{\textrm{labelings}} \prod_{\text{edges}} d_e \prod_{\text{tetrahedra}} \frac{F^{ijm}_{kln}}{\sqrt{d_m d_n}}
 \enspace 
\end{equation}
if~$\cC$ is multiplicity-free.
Here, the sum is over edge labelings of the triangulation by particles from the category~$\cC$.  The index $e$ on $d$ is the label of an edge, while the indices $i, \ldots, n$ are the labels of the six edges involved in a tetrahedron, ordered and oriented following certain rules.  The fusion tensor $F$, the quantum dimensions~$d$ and the total quantum dimension $\cD$ are parameters of~$\cC$.  $\abs V$ is the number of vertices of the triangulation.  The topological invariance of $\TV_\cC(M)$ follows from the fact that any two triangulations of $M$ can be related by a finite sequence of local Pachner moves~\cite{Pachner91}, under which the above quantity is invariant.  In this paper we consider multiplicity-free unitary modular tensor categories, which include the $\SU(2)_k$ case, but are not as general as spherical tensor categories.  

\begin{figure}
\includegraphics[scale=1]{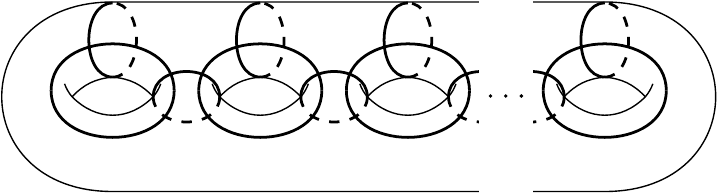}
\caption{A Dehn twist is a $2 \pi$ rotation about a closed curve.  The Dehn twists about the $3g-1$ curves shown above generate the full mapping class group of the genus-$g$ surface~\cite{Lickorish64generators}.}  
\label{fig:Dehn}
\end{figure}

To formulate a BQP-complete problem~\cite{WocjanZhang06BQPcomplete} of estimating the Turaev-Viro invariant, we require a presentation of $3$-manifolds known as a Heegaard splitting.  Consider two genus-$g$ handlebodies (e.g., the solid torus for $g=1$).  They can be glued together, to give a $3$-manifold, using a self-homeomorphism of the genus-$g$ surface.  The set of orientation-preserving self-homeomorphisms modulo those isotopic to the identity form the mapping class group $\MCG(g)$ of the surface.  It is an infinite group generated by the $3g-1$ Dehn twists illustrated in \figref{fig:Dehn}.  A Heegaard splitting thus consists of a natural number $g$ and an element $x \in \MCG(g)$, defining a manifold $M(g, x)$.  Every compact, orientable $3$-manifold can be obtained in this way, up to homeomorphism.  

\begin{theorem} \label{t:TVproblem}
For any fixed multiplicity-free unitary modular tensor category $\cC$, there is a quantum algorithm that, given $\delta, \epsilon > 0$, $g \in \N$ and a length-$m$ word $x$ in the Dehn-twist generators of $\MCG(g)$ from \figref{fig:Dehn}, runs in time $\mathrm{poly}(g, m, \log 1/\delta, 1/\epsilon)$ and, except with probability at most $\delta$, outputs an approximation of $\TV_\cC(M(g, x))$ to within~$\pm \cD^{2(g-1)}\, \epsilon$.  

Conversely, for $\cC$ the category associated to
 $\SU(2)_k$ or $\SO(3)_k$ for $k\geq 3$ such that $k+2$~is prime, it is BQP-hard to decide whether $\cD^{2(1-g)} \, \TV_\cC(M(g, x))$ is greater than $2/3$ or less than $1/3$.  
More precisely, given any quantum circuit $\Upsilon$ of $T$ two-qubit gates acting on $n$ qubits $\ket{0^n}$, with output either $0$ or $1$, 
one can classically find in polynomial time a 
word $x = x_1 \ldots x_m$ in the standard Dehn-twist generators of~$\MCG(g)$, with $g = n + 1$ and $m = \mathrm{poly}(T)$, such that 
\begin{equation}
\big\lvert \Pr[\text{$\Upsilon$ outputs $1$}] - \cD^{2(1-g)} \, \TV_\cC(M(g, x)) \big\rvert < 1/6
 \enspace .
\end{equation}
\end{theorem}

The additive approximation error is exponential in~$g$.  
Complexity-theoretic reasons make it unlikely that quantum computers can efficiently obtain a multiplicative or otherwise presentation-independent error~\cite{Kuperberg09}.  

In fact, a similar statement to \thmref{t:TVproblem} also holds for approximating the Witten-Reshetikhin-Turaev (WRT) invariants~\cite{Witten89, ReshetikhinTuraev91}.  For any $g$, a modular category~$\cC$ can be used to define a projective representation $\rho_{\cC, g}: \MCG(g) \rightarrow \GL(\cH_{\cC, g})$.  This representation will be given below.  The WRT invariant for a $3$-manifold~$M(g, x)$ is then given by a matrix element 
\begin{equation} \label{eq:wrtinvariant}
\WRT_\cC(M(g,x)) = \cD^{g-1} \bra{v_{\cC, g}} \rho_{\cC, g}(x) \ket{v_{\cC, g}}\ ,
\end{equation}
where $\ket{v_{\cC, g}} \in \cH_{\cC, g}$ is a certain unit-normalized
vector.  As the representation is projective,~$\WRT_\cC$ is a
3-manifold invariant only up to a multiple of $e^{2\pi i c/24}$ where $c$
is called the central charge.
(Eq.~\eqref{eq:wrtinvariant} is the Crane-Kohno-Kontsevich presentation~\cite{Crane91, Kohno92, Kontsevich88} of the WRT invariant, which is more commonly defined in terms of a Dehn surgery presentation of~$M$.  Equivalence of these definitions for $\cC = \SU(2)_k$ is shown in~\cite{Piunikhin93}; see also~\cite[Sec.~2.4]{Kohno02}.)  

The fact that Eq.~\eqnref{eq:wrtinvariant} indeed gives an invariant can be established by studying the problem of when two Heegaard splittings~$(g, x)$ and $(g', x')$ describe homeomorphic manifolds.  Since taking the connected sum of a manifold $M$ with the $3$-sphere~$S^3$ does not change the manifold, i.e., $M\# S^3\cong M$, the standard Heegaard splitting of~$S^3$ into two genus-one handlebodies allows defining a ``stabilization'' map $(g, x) \mapsto (g+1, \tilde x)$ such that $M(g, x) \cong M(g+1, \tilde x)$.  A general theorem of Reidemeister~\cite{Reidemeister33} and Singer~\cite{Singer33} asserts that $M(g, x) \cong M(g', x')$ if and only if $(g,x)$ and $(g',x')$ are equivalent under stabilization and the following algebraic equivalence relation for the case of equal genus~\cite{Funar95} 
\begin{equation}
(g, x) \equiv (g, x') \; \textrm{ if $x = yx'z$ with $y, z \in \MCG^+(g)$}
 \enspace .
\end{equation}
Here $\MCG^+(g) \subset \MCG(g)$ is the subgroup of self-homeomorphisms (classes) of the genus-$g$ surface that extend to 
the genus-$g$ handlebody.  Invariance of $\WRT_\cC(M(g,x))$ now follows essentially from the fact that $\ket{v_{\cC, g}}$ is invariant under the action of~$\MCG^+(g)$.  

The Turaev-Viro and WRT invariants are related by   
\begin{equation} \label{eq:tvwrtinvariant}
\TV_\cC(M) = \abs{\WRT_\cC(M)}^2
\end{equation}
as shown by Turaev~\cite{Turaev91} and Walker~\cite{Walker91} (see also~\cite{Turaev94book, Roberts95}).  
In~\cite{KoenigKuperbergReichardt10TVcode}, Eq.~\eqnref{eq:tvwrtinvariant} is discussed in the category-theoretic formalism used here.
Identities~\eqnref{eq:wrtinvariant} and~\eqnref{eq:tvwrtinvariant}, together with density and locality properties of the representations~$\rho_{\cC, g}$, are the basis of our BQP-completeness proof.  

Previously, a quantum algorithm for approximating the $\SU(2)_k$ Turaev-Viro and WRT invariants was given by Garnerone \emph{et al.}~\cite{GarneroneMarzuoliRasetti07}, assuming the manifold is specified by Dehn surgery rather than a Heegaard splitting.  BQP-hardness of the approximation was left as an open problem.  In unpublished work, Bravyi and Kitaev have proven the BQP-completeness of the problem of approximating the $\SU(2)_4$ WRT invariant of $3$-manifolds with boundary~\cite{BravyiKitaev00}, where the manifold is specified using Morse functions.  We remark that one can use Arad and Landau's quantum algorithm for approximating tensor network contractions to compute the Turaev-Viro invariant of a triangulated manifold~\cite{AradLandau08tensor}.  While this algorithm would run polynomially in the number of tetrahedra, its precision depends on the order in which tensors are contracted and may be trivial.  

We will only briefly describe the space $\cH_{\cC, g}$, the representation $\rho_{\cC, g} : \MCG(g) \rightarrow \GL(\cH_{\cC, g})$ and the state $\ket{v_{\cC, g}} \in \cH_{\cC, g}$ from Eq.~\eqnref{eq:wrtinvariant}.  Details are in~\cite{Crane91, Kohno92, Kontsevich88, KoenigKuperbergReichardt10TVcode}.  

Let $\cC$ be a multiplicity-free unitary modular tensor category.  It specifies a set of particles $i$ with quantum dimensions $d_i > 0$, and including a trivial particle~$\vac$.  The total quantum dimension is $\cD = \sqrt{\sum_i d_i^2}$.  $\cC$ additionally specifies a particle duality map $i \mapsto i^*$, fusion rules, $F$-symbols $F^{ijm}_{kln}$ and $R$-symbols $R_i^{jk}$.  These tensors obey certain identities, such as the pentagon and hexagon equations, which can be found in, e.g., \cite{Preskill98notes, KoenigKuperbergReichardt10TVcode}. 

\begin{figure}
\includegraphics[scale=.42,angle=0]{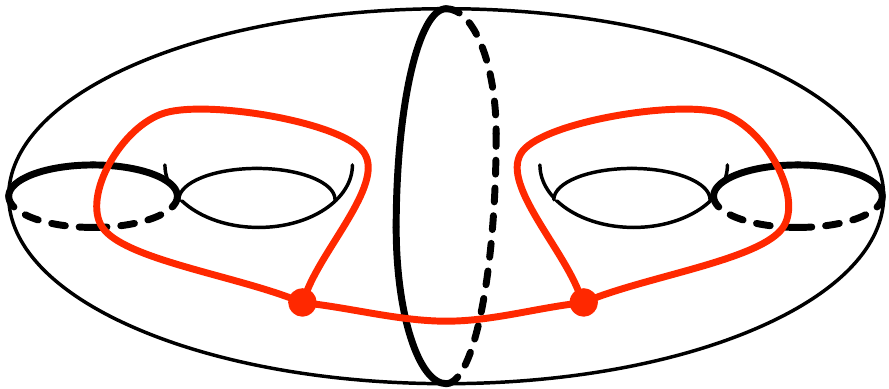} \includegraphics[scale=.42,angle=0]{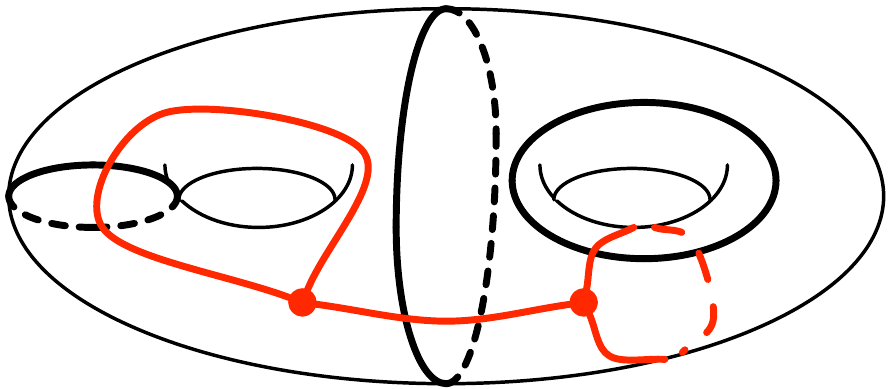} \includegraphics[scale=.42,angle=0]{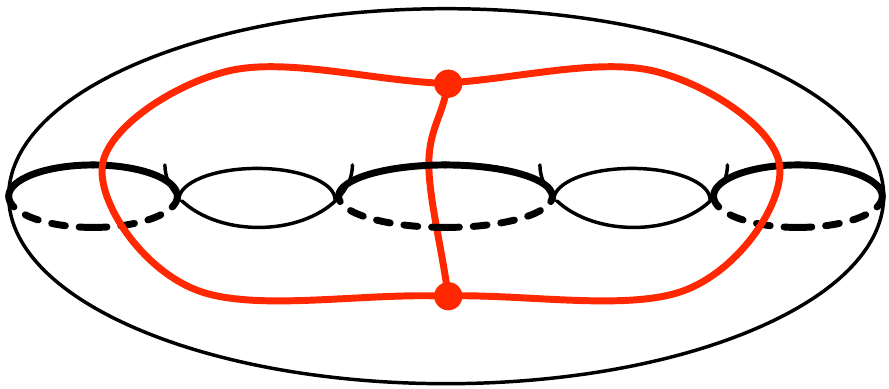}
\caption{Three examples of decompositions of the genus-two surface $\Sigma_2$ into three-punctured spheres.  In each case, a trivalent adjacency graph of the punctured spheres is shown in red.}  \label{fig:dualgraphexamples}
\end{figure}

Let $g \in \N$, $g \geq 2$.  The space $\cH_{\cC, g}$ can be defined by specifying an orthonormal basis.  Decompose the genus-$g$ surface~$\Sigma_g$ into three-punctured spheres (or ``pants'') by cutting along $3g - 3$ noncontractible curves, as illustrated in \figref{fig:dualgraphexamples}.  Dual to such a decomposition is a trivalent graph~$\Gamma$.  Direct arbitrarily the edges of $\Gamma$.  A basis vector $\ket \ell_\Gamma$ is a fusion-consistent labeling of the edges of $\Gamma$ by particles of the category~$\cC$.  Fusion-consistency is defined by the fusion rules, i.e., a set of triples~$(i,j,k)$ that are allowed to meet at every vertex, and particle duality, which switches the direction of an edge, replacing a label~$i$ by the antiparticle~$i^*$.  Define the states $\cB_{\Gamma}:=\{\ket{\ell}_\Gamma\}_{\ell}$ to be orthonormal, and their span to be $\cH_{\cC, g}$.  Note that this
definition gives a natural encoding of $\cH_{\cC, g}$ into qudits, with one qudit to store the label of each edge of~$\Gamma$.  The directed graph $\Gamma$ can be stored in a classical register.  

The above definition depends on~$\Gamma$, but alternative pants decompositions simply represent different 
bases~$\cB_\Gamma$ for the same Hilbert space.  To convert between all possible pants decompositions of~$\Sigma_g$ we need two moves, each corresponding to a local unitary operator.  

The $F$ move relates bases that differ by a ``flip" of a cut between two three-punctured spheres.  In the qudit encoding, it is a five-qudit unitary, with four control qudits.  Its action is given by
\begin{equation} \label{eq:Fmove}
\raisebox{-1.3cm}{\includegraphics[scale=.5]{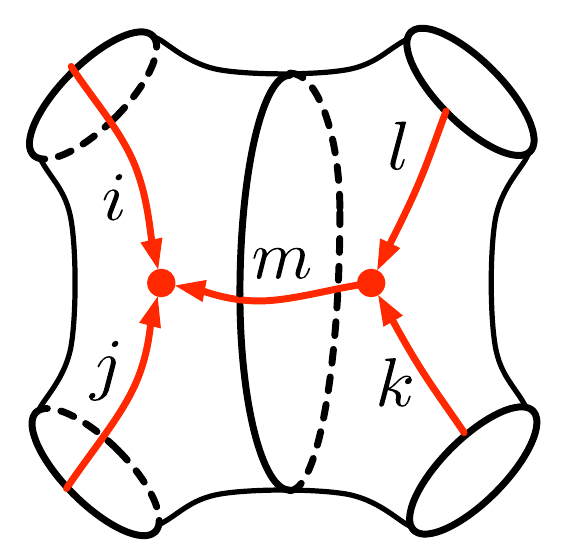}}
=
\sum_n F^{i j m}_{k l n} 
\raisebox{-1.3cm}{\includegraphics[scale=.5]{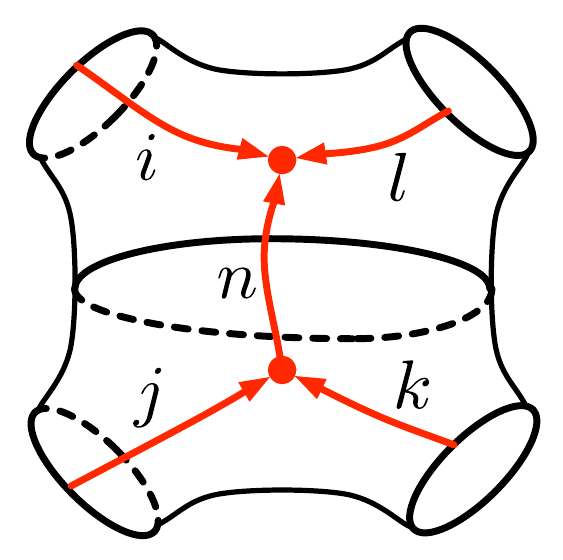}}
\end{equation}

The $S$ move applies when two boundaries of a single three-punctured sphere are connected.  It is a two-qudit unitary, with one control qudit, and its action is given by 
\begin{equation} \label{eq:Smove}
\raisebox{-1.3cm}{\includegraphics[scale=.5]{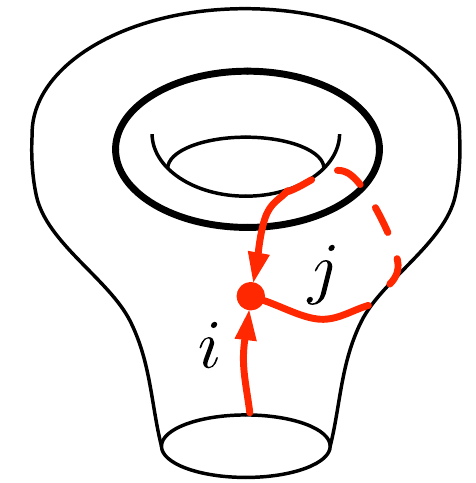}}
=
\sum_k S^i_{jk} 
\raisebox{-1.3cm}{\includegraphics[scale=.5]{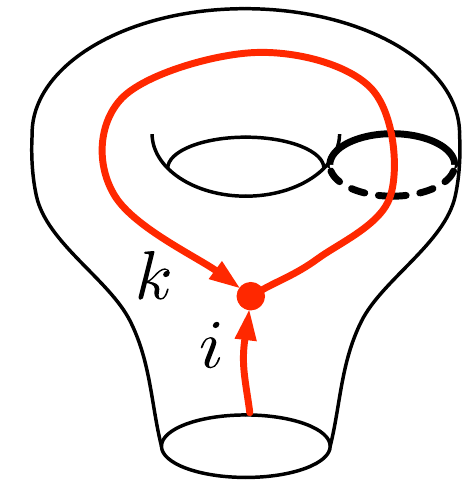}}
\end{equation}
Most presentations of modular tensor categories do not explicitly provide values for $S^i_{jk}$.  However, as discussed in~\cite{Walker91}, $S^i_{jk}$~can be calculated by the identity 
\begin{equation}
\cD S^i_{jk} 
= 
\sum_{\substack{l:\,(j,k^*,\ell)\\~\textrm{fusion-consistent}}} \hspace{-0.5cm}
F^{i k^* k}_{l j^* j} \frac{d_l}{\sqrt {d_i}} R^{k j^*}_l R^{j k^*}_{l^*} 
=
\raisebox{-.55cm}{\includegraphics[scale=1]{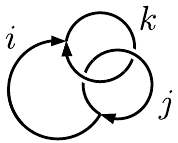}} 
\end{equation}
(The last expression uses ribbon graph notation.)  

The action~$\rho_{\cC, g}$ of~$\MCG(g)$ on~$\cH_{\cC, g}$ can now be specified by the action of the Dehn-twist generators on basis vectors.  For a Dehn twist about a curve $\sigma$, apply a sequence of $F$ and $S$ moves to change into a basis $\cB_\Gamma$, i.e., a pants decomposition of $\Sigma_g$, in which $\sigma$ divides two three-punctured spheres.  In such a basis, the Dehn twist acts diagonally: if the edge of~$\Gamma$ crossing $\sigma$ has label~$i$, the twist applies a phase shift of~$R^{ii^*}_0$.  

To complete the definition of~$\WRT_\cC(M(g,x))$ from Eq.~\eqnref{eq:wrtinvariant}, it remains to define the state~$\ket{v_{\cC, g}}$.  As on the right-hand side of Eq.~\eqnref{eq:Smove}, decompose $\Sigma_g$ with a meridional cut through each handle.  Then $\ket{v_{\cC, g}}$ is the state in which every edge of~$\Gamma$ is labeled by~$\vac$, the trivial particle.  

Let us now prove \thmref{t:TVproblem}.  Although not obvious from Eq.~\eqnref{eq:triang_invar}, the original tensor-network-contraction-based definition of the Turaev-Viro invariant, \thmref{t:TVproblem} is a straightforward consequence of the definition based on the representation~$\rho_{\cC, g}$, and of known density results.  

\begin{figure}
\subfigure[]{\label{fig:encoding}\raisebox{0in}{\includegraphics{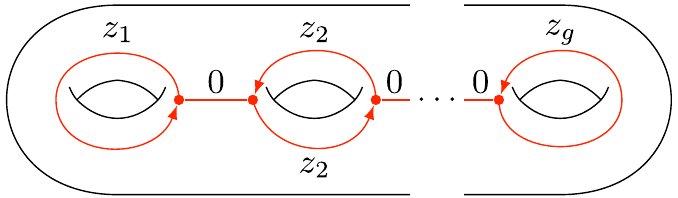}}}\\\subfigure[]{\label{fig:local}\raisebox{0in}{\includegraphics{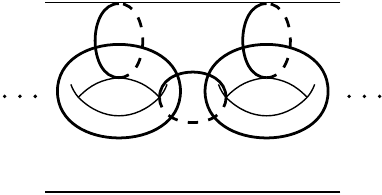}}}
\caption{(a) A $g$-qubit state $\ket z$, $z \in \{0,1\}^g$, can be encoded into $\cH_{\cC,g}$ for the genus-$g$ handlebody.  (b) Any two-qubit gate can be approximated within the codespace using the Dehn twists involving the two corresponding handles.} 
\end{figure}

The Turaev-Viro and WRT invariants for $M(g, x)$ can be approximated essentially by implementing $\rho_{\cC, g}(x)$.  The algorithm maintains a classical register storing the graph~$\Gamma$, together with a quantum register containing the current state in~$\cH_{\cC, g}$ in the basis $\cB_\Gamma$.  If~$\cC$ has~$N$ particle types, the algorithm uses an $N$-dimensional qudit for each edge of~$\Gamma$.  Then~$\rho_{\cC, g}(x_j)$ can be applied by using a sequence of $F$ and $S$ moves, i.e., certain local unitaries, to change to a basis in which $x_j$ acts diagonally.  Since $x_j$~is one of the generators from \figref{fig:Dehn}, starting with the graph $\Gamma$ of \figref{fig:encoding} (for which every edge is labeled~$\vac$ in $\ket{v_{\cC, g}}$) at most one $F$ and one $S$ move are needed.  An estimate to within $\epsilon$ of the desired matrix element $\bra{v_{\cC, g}} \rho_{\cC, g}(x) \ket{v_{\cC, g}}$ can be given, except with probability~$\delta$, using $O(\log (1/\delta) /\epsilon^2)$ Hadamard tests, as in~\cite{AharonovJonesLandau06Jones}.  

\def\circuit{\Upsilon}
To prove BQP-hardness we reduce from the BQP-complete problem of deciding whether $\lvert \bra{0^g} \circuit \ket{0^g}\rvert^2$ is larger than $5/6$ or less than $1/6$, given the $g$-qubit quantum circuit~$\circuit$~\cite{AharonovJonesLandau06Jones}.  Let~$\cC$ be the modular tensor category associated with~$\SU(2)_k$ or $\SO(3)_k$, with $k \geq 3$ and $k+2$ prime.  Given~$\circuit$ consisting of $T$ two-qubit gates, our aim is to construct efficiently the Heegaard splitting~$(g,x)$ of a manifold~$M=M(g,x)$ such that $\cD^{2(1-g)} \TV_\cC(M)$ approximates $\lvert \bra{0^g} \circuit \ket{0^g}\rvert^2$.  As illustrated in \figref{fig:encoding}, we use one handle of a genus-$g$ handlebody to encode each qubit.  Such a labeling is fusion-consistent, and the encoding of the initial state $\ket{0^g}$ is exactly $\ket{v_{\cC, g}} \in \cH_{\cC, g}$.   As shown in~\cite{FLW_dense, LarsenWang05dense}, for $\cC = \SO(3)_k$ the representation $\rho_{\cC, g}$ has a dense image, up to phases, in the group of unitaries on~$\cH_{\cC,g}$, for $g \geq 2$.  By the density for $g = 2$ and the Solovay-Kitaev theorem~\cite{NielsenChuang00}, therefore any two-qubit gate can be approximated in the codespace to precision $1/(6T)$ by applying a $(\log T)^{O(1)}$-long sequence of the five Dehn twists shown in \figref{fig:local}.  This holds also for $\cC = \SU(2)_k$, as $\SO(3)_k$ is just the restriction of $\SU(2)_k$ to particles with integer spins.  Thus we obtain a polynomial-length word~$x = x_1\cdots x_{poly(T)}$ in the Dehn-twist generators whose action approximates $\circuit$ on the codespace.  Then $\bra{v_{\cC, g}} \rho_{\cC, g}(M(g, x)) \ket{v_{\cC, g}}$ approximates~$\bra{0^g} \circuit \ket{0^g}$.  

This work demonstrates how quantum physics, in the form of TQFTs, can inspire new quantum algorithms for problems based on topology and tensor networks.  The approach taken here realizes in a sense the traditional vision of quantum computers as universal simulators for physical systems, but with a different outcome: it provides a purely mathematical problem whose difficulty exactly captures the power of a quantum computer.  

\bigskip
S.J.\ acknowledges support from the Sherman Fairchild Foundation and NSF grant PHY-0803371.  R.K.\ acknowledges support by the Swiss National Science Foundation (SNF) under grant PA00P2-126220.  B.R.\ and G.A.\ acknowledge support from NSERC and  ARO.  Some of this research was conducted at the Kavli Institute for Theoretical Physics, supported by NSF grant PHY05-51164.

\bibliographystyle{apsrev}
\bibliography{simple_tv}

\begin{thebibliography}{34}
\expandafter\ifx\csname natexlab\endcsname\relax\def\natexlab#1{#1}\fi
\expandafter\ifx\csname bibnamefont\endcsname\relax
  \def\bibnamefont#1{#1}\fi
\expandafter\ifx\csname bibfnamefont\endcsname\relax
  \def\bibfnamefont#1{#1}\fi
\expandafter\ifx\csname citenamefont\endcsname\relax
  \def\citenamefont#1{#1}\fi
\expandafter\ifx\csname url\endcsname\relax
  \def\url#1{\texttt{#1}}\fi
\expandafter\ifx\csname urlprefix\endcsname\relax\def\urlprefix{URL }\fi
\providecommand{\bibinfo}[2]{#2}
\providecommand{\eprint}[2][]{\url{#2}}

\bibitem[{\citenamefont{Freedman
  et~al.}(2002{\natexlab{a}})\citenamefont{Freedman, Kitaev, and
  Wang}}]{FreedmanKitaevWang00}
\bibinfo{author}{\bibfnamefont{M.~H.} \bibnamefont{Freedman}},
  \bibinfo{author}{\bibfnamefont{A.~Yu.} \bibnamefont{Kitaev}},
  \bibnamefont{and} \bibinfo{author}{\bibfnamefont{Z.}~\bibnamefont{Wang}},
  \bibinfo{journal}{Comm. Math. Phys.} \textbf{\bibinfo{volume}{227}},
  \bibinfo{pages}{587} (\bibinfo{year}{2002}{\natexlab{a}}).

\bibitem[{\citenamefont{Walker}(1991)}]{Walker91}
\bibinfo{author}{\bibfnamefont{K.}~\bibnamefont{Walker}}, On {W}itten's 3-manifold invariants
  (\bibinfo{year}{1991}), \bibinfo{note}{\href{http://canyon23.net/math/}{http://canyon23.net/math/}}.

\bibitem[{\citenamefont{Aharonov et~al.}(2006)\citenamefont{Aharonov, Jones,
  and Landau}}]{AharonovJonesLandau06Jones}
\bibinfo{author}{\bibfnamefont{D.}~\bibnamefont{Aharonov}},
  \bibinfo{author}{\bibfnamefont{V.}~\bibnamefont{Jones}}, \bibnamefont{and}
  \bibinfo{author}{\bibfnamefont{Z.}~\bibnamefont{Landau}}, in
  \emph{\bibinfo{booktitle}{Proc. 38th ACM STOC}} (\bibinfo{year}{2006}), pp.
  \bibinfo{pages}{427--436}, \eprint{arXiv:quant-ph/0511096}.

\bibitem[{\citenamefont{Aharonov and Arad}(2006)}]{AharonovArad06Jones}
\bibinfo{author}{\bibfnamefont{D.}~\bibnamefont{Aharonov}} \bibnamefont{and}
  \bibinfo{author}{\bibfnamefont{I.}~\bibnamefont{Arad}}
  (\bibinfo{year}{2006}), \eprint{arXiv:quant-ph/0605181}.

\bibitem[{\citenamefont{Wocjan and Yard}(2007)}]{WocjanYard07Jones}
\bibinfo{author}{\bibfnamefont{P.}~\bibnamefont{Wocjan}} \bibnamefont{and} \bibinfo{author}{\bibfnamefont{J.}~\bibnamefont{Yard}}, 
\bibinfo{journal}{Quantum Inf. Comput.} \textbf{\bibinfo{volume}{8}},
  \bibinfo{pages}{147} (\bibinfo{year}{2008}{\natexlab{a}}). 

\bibitem[{\citenamefont{Garnerone et~al.}(2006)\citenamefont{Garnerone,
  Marzuoli, and Rasetti}}]{GarneroneMarzuoliRasetti06Jones}
\bibinfo{author}{\bibfnamefont{M.}~\bibnamefont{Rasetti}},
  \bibinfo{author}{\bibfnamefont{S.}~\bibnamefont{Garnerone}}, \bibnamefont{and}
  \bibinfo{author}{\bibfnamefont{A.}~\bibnamefont{Marzuoli}}, 
\bibinfo{journal}{Int. J. Quantum Inf.} \textbf{\bibinfo{volume}{6}},
  \bibinfo{pages}{773} (\bibinfo{year}{2008}{\natexlab{a}}). 

\bibitem[{\citenamefont{Turaev and Viro}(1992)}]{TuraevViro92}
\bibinfo{author}{\bibfnamefont{V.~G.} \bibnamefont{Turaev}} \bibnamefont{and}
  \bibinfo{author}{\bibfnamefont{O.~Y.} \bibnamefont{Viro}},
  \bibinfo{journal}{Topology} \textbf{\bibinfo{volume}{31}},
  \bibinfo{pages}{865} (\bibinfo{year}{1992}).

\bibitem[{\citenamefont{Barrett and
  Westbury}(1996)}]{BarrettWestbury96invariants}
\bibinfo{author}{\bibfnamefont{J.~W.} \bibnamefont{Barrett}} \bibnamefont{and}
  \bibinfo{author}{\bibfnamefont{B.~W.} \bibnamefont{Westbury}},
  \bibinfo{journal}{Trans. Amer. Math. Soc.} \textbf{\bibinfo{volume}{348}},
  \bibinfo{pages}{3997} (\bibinfo{year}{1996}). 

\bibitem[{\citenamefont{Moise}(1952)}]{Moise52}
\bibinfo{author}{\bibfnamefont{E.~E.} \bibnamefont{Moise}},
  \bibinfo{journal}{Ann. Math. (2)} \textbf{\bibinfo{volume}{56}},
  \bibinfo{pages}{96} (\bibinfo{year}{1952}).

\bibitem[{\citenamefont{Pachner}(1991)}]{Pachner91}
\bibinfo{author}{\bibfnamefont{U.}~\bibnamefont{Pachner}},
  \bibinfo{journal}{Eur. J. Combin.} \textbf{\bibinfo{volume}{12}},
  \bibinfo{pages}{129} (\bibinfo{year}{1991}).

\bibitem[{\citenamefont{Wocjan and Zhang}(2006)}]{WocjanZhang06BQPcomplete}
\bibinfo{author}{\bibfnamefont{P.}~\bibnamefont{Wocjan}} \bibnamefont{and}
  \bibinfo{author}{\bibfnamefont{S.}~\bibnamefont{Zhang}}
  (\bibinfo{year}{2006}), \eprint{arXiv:quant-ph/0606179}.

\bibitem[{\citenamefont{Lickorish}(1964)}]{Lickorish64generators}
\bibinfo{author}{\bibfnamefont{W.~B.~R.} \bibnamefont{Lickorish}},
  \bibinfo{journal}{Proc. Cambridge Philos. Soc.}
  \textbf{\bibinfo{volume}{60}}, \bibinfo{pages}{769} (\bibinfo{year}{1964}),
  \bibinfo{note}{continued in vol. 62, pg. 679-681 (1966)}.

\bibitem[{\citenamefont{Kuperberg}(2009)}]{Kuperberg09}
\bibinfo{author}{\bibfnamefont{G.}~\bibnamefont{Kuperberg}} (\bibinfo{year}{2009}),
  \bibinfo{journal}{arXiv:0908.0512}.

\bibitem[{\citenamefont{Witten}(1989)}]{Witten89}
\bibinfo{author}{\bibfnamefont{E.}~\bibnamefont{Witten}},
  \bibinfo{journal}{Comm. Math. Phys.} \textbf{\bibinfo{volume}{121}},
  \bibinfo{pages}{351} (\bibinfo{year}{1989}).

\bibitem[{\citenamefont{Reshetikhin and Turaev}(1991)}]{ReshetikhinTuraev91}
\bibinfo{author}{\bibfnamefont{N.}~\bibnamefont{Reshetikhin}} \bibnamefont{and}
  \bibinfo{author}{\bibfnamefont{V.~G.} \bibnamefont{Turaev}},
  \bibinfo{journal}{Invent. Math.} \textbf{\bibinfo{volume}{103}}
  (\bibinfo{year}{1991}).

\bibitem[{\citenamefont{Crane}(1991)}]{Crane91}
\bibinfo{author}{\bibfnamefont{L.}~\bibnamefont{Crane}},
  \bibinfo{journal}{Comm. Math. Phys.} \textbf{\bibinfo{volume}{135}},
  \bibinfo{pages}{615} (\bibinfo{year}{1991}).

\bibitem[{\citenamefont{Kohno}(1992)}]{Kohno92}
\bibinfo{author}{\bibfnamefont{T.}~\bibnamefont{Kohno}},
  \bibinfo{journal}{Topology} \textbf{\bibinfo{volume}{31}},
  \bibinfo{pages}{203} (\bibinfo{year}{1992}).

\bibitem[{\citenamefont{Kontsevich}(1988)}]{Kontsevich88}
\bibinfo{author}{\bibfnamefont{M.}~\bibnamefont{Kontsevich}}
  (\bibinfo{year}{1988}), \bibinfo{note}{preprint of the Centre de Physique
  Theorique Marseille, CPT-88/p2189}.

\bibitem[{\citenamefont{Piunikhin}(1993)}]{Piunikhin93}
\bibinfo{author}{\bibfnamefont{S.}~\bibnamefont{Piunikhin}},
  \bibinfo{journal}{J. Knot Theor. Ramif.} \textbf{\bibinfo{volume}{2}},
  \bibinfo{pages}{65} (\bibinfo{year}{1993}).

\bibitem[{\citenamefont{Kohno}(2002)}]{Kohno02}
\bibinfo{author}{\bibfnamefont{T.}~\bibnamefont{Kohno}},
  \emph{\bibinfo{title}{Conformal Field Theory and Topology}}
  (\bibinfo{publisher}{American Mathematical Society}, \bibinfo{year}{2002}),
  vol. \bibinfo{volume}{210} of \emph{\bibinfo{series}{Translations of
  Mathematical Monographs}}, chap. \bibinfo{chapter}{2.4}.

\bibitem[{\citenamefont{Reidemeister}(1933)}]{Reidemeister33}
\bibinfo{author}{\bibfnamefont{K.}~\bibnamefont{Reidemeister}},
  \emph{\bibinfo{title}{Zur dreidimensionalen {T}opologie}}, Abh. Math. Sem.
  (\bibinfo{publisher}{Univ. Hamburg}, \bibinfo{year}{1933}).

\bibitem[{\citenamefont{Singer}(1933)}]{Singer33}
\bibinfo{author}{\bibfnamefont{J.}~\bibnamefont{Singer}},
  \bibinfo{journal}{Trans. Amer. Math. Soc.} \textbf{\bibinfo{volume}{35}},
  \bibinfo{pages}{88} (\bibinfo{year}{1933}), ISSN \bibinfo{issn}{00029947},
 \url{http://www.jstor.org/stable/1989314}.

\bibitem[{\citenamefont{Funar}(1995)}]{Funar95}
\bibinfo{author}{\bibfnamefont{L.}~\bibnamefont{Funar}},
  \bibinfo{journal}{Comm. Math. Phys.} \textbf{\bibinfo{volume}{171}},
  \bibinfo{pages}{405} (\bibinfo{year}{1995}).

\bibitem[{\citenamefont{Turaev}(1991)}]{Turaev91}
\bibinfo{author}{\bibfnamefont{V.~G.} \bibnamefont{Turaev}}
 \emph{\bibinfo{title}{Topology of shadows}} 
  (\bibinfo{year}{1991}), \bibinfo{note}{preprint}.

\bibitem[{\citenamefont{Turaev}(1994)}]{Turaev94book}
\bibinfo{author}{\bibfnamefont{V.~G.} \bibnamefont{Turaev}},
  \emph{\bibinfo{title}{Quantum invariants of knots and 3-manifolds}},
  vol.~\bibinfo{volume}{18} of \emph{\bibinfo{series}{de Gruyter studies in
  mathematics}} (\bibinfo{publisher}{de Gruyter}, \bibinfo{address}{New York},
  \bibinfo{year}{1994}).

\bibitem[{\citenamefont{Roberts}(1995)}]{Roberts95}
\bibinfo{author}{\bibfnamefont{J.~D.} \bibnamefont{Roberts}},
  \bibinfo{journal}{Topology} \textbf{\bibinfo{volume}{34}},
  \bibinfo{pages}{771} (\bibinfo{year}{1995}).

\bibitem[{\citenamefont{K{\"o}nig et~al.}(2010)\citenamefont{K{\"o}nig,
  Kuperberg, and Reichardt}}]{KoenigKuperbergReichardt10TVcode}
\bibinfo{author}{\bibfnamefont{R.}~\bibnamefont{K{\"o}nig}},
  \bibinfo{author}{\bibfnamefont{G.}~\bibnamefont{Kuperberg}},
  \bibnamefont{and} \bibinfo{author}{\bibfnamefont{B.~W.}
  \bibnamefont{Reichardt}} (\bibinfo{year}{2010}), \eprint{arXiv:1002.2816 [quant-ph]}.  

\bibitem[{\citenamefont{Garnerone et~al.}(2007)\citenamefont{Garnerone,
  Marzuoli, and Rasetti}}]{GarneroneMarzuoliRasetti07}
\bibinfo{author}{\bibfnamefont{S.}~\bibnamefont{Garnerone}},
  \bibinfo{author}{\bibfnamefont{A.}~\bibnamefont{Marzuoli}}, \bibnamefont{and}
  \bibinfo{author}{\bibfnamefont{M.}~\bibnamefont{Rasetti}}
  (\bibinfo{year}{2007}), \eprint{arXiv:quant-ph/0703037}.

\bibitem[{\citenamefont{Bravyi and Kitaev}(2000)}]{BravyiKitaev00}
\bibinfo{author}{\bibfnamefont{S.~B.} \bibnamefont{Bravyi}} \bibnamefont{and}
  \bibinfo{author}{\bibfnamefont{A.~Yu.} \bibnamefont{Kitaev}}
\emph{\bibinfo{title}{Quantum invariants of 3-manifolds and quantum computation}}
  (\bibinfo{year}{2000}), \bibinfo{note}{unpublished}.

\bibitem[{\citenamefont{Arad and Landau}(2008)}]{AradLandau08tensor}
\bibinfo{author}{\bibfnamefont{I.}~\bibnamefont{Arad}} \bibnamefont{and}
  \bibinfo{author}{\bibfnamefont{Z.}~\bibnamefont{Landau}}
  (\bibinfo{year}{2008}), \eprint{arXiv:0805.0040}.

\bibitem[{\citenamefont{Preskill}(1998)}]{Preskill98notes}
\bibinfo{author}{\bibfnamefont{J.}~\bibnamefont{Preskill}},
  \emph{\bibinfo{title}{Physics 219 lecture notes}}, \bibinfo{howpublished}{\url{http://theory.caltech.edu/~preskill/ph229/}} (\bibinfo{year}{1998}).

\bibitem[{\citenamefont{Freedman
  et~al.}(2002{\natexlab{b}})\citenamefont{Freedman, Larsen, and
  Wang}}]{FLW_dense}
\bibinfo{author}{\bibfnamefont{M.~H.} \bibnamefont{Freedman}},
  \bibinfo{author}{\bibfnamefont{M.~J.} \bibnamefont{Larsen}},
  \bibnamefont{and} \bibinfo{author}{\bibfnamefont{Z.}~\bibnamefont{Wang}},
  \bibinfo{journal}{Comm. Math. Phys.} \textbf{\bibinfo{volume}{228}},
  \bibinfo{pages}{177} (\bibinfo{year}{2002}{\natexlab{b}}).

\bibitem[{\citenamefont{Larsen and Wang}(2005)}]{LarsenWang05dense}
\bibinfo{author}{\bibfnamefont{M.}~\bibnamefont{Larsen}} \bibnamefont{and}
  \bibinfo{author}{\bibfnamefont{Z.}~\bibnamefont{Wang}},
  \bibinfo{journal}{Comm. Math. Phys.} \textbf{\bibinfo{volume}{260}},
  \bibinfo{pages}{641} (\bibinfo{year}{2005}).

\bibitem[{\citenamefont{Nielsen and Chuang}(2000)}]{NielsenChuang00}
\bibinfo{author}{\bibfnamefont{M.}~\bibnamefont{Nielsen}} \bibnamefont{and}
  \bibinfo{author}{\bibfnamefont{I.}~\bibnamefont{Chuang}},
  \emph{\bibinfo{title}{Quantum Computation and Quantum Information}}
  (\bibinfo{publisher}{Cambridge University Press}, \bibinfo{year}{2000}).

\end{thebibliography}

\end{document}